\documentclass{myws}

\begin{document}
\def\epem{\mbox{e}^+\mbox{e}^-}
\def\eegg{\epem\to\gamma\gamma}
\def\eeggg{\epem\to\gamma\gamma(\gamma)}
\def\ggg{\gamma\gamma(\gamma)}
\def\ct{\cos{\theta}}
\def\cte{\cos{\theta^{\ast}}}
\def\xmc{\left(\frac{d\sigma}{d\Omega}\right)_{\rm MC}}
\def\xb{\left(\frac{d\sigma}{d\Omega}\right)_{\rm Born}}
\def\xl{\left(\frac{d\sigma}{d\Omega}\right)_{\Lambda_{\pm}}}
\def\xq{\left(\frac{d\sigma}{d\Omega}\right)_{\rm \Lambda '}}
\def\xe{\left(\frac{d\sigma}{d\Omega}\right)_{\rm e^{\ast}}}
\def\xg{\left(\frac{d\sigma}{d\Omega}\right)_{\rm M_S}}
\def\xsn{\frac{d\sigma}{d\Omega}}
\def\be{\begin{equation}}
\def\ee{\end{equation}}

\title{TESTS OF QED WITH MULTI-PHOTONIC FINAL STATES}

\author{Kirsten Sachs}

\address{Carleton University, 
1125 Colonel By Drive, Ottawa, ON K15 5B6, Canada\\ 
E-mail: Kirsten.Sachs@cern.ch}



\maketitle

\abstracts{
In the Standard Model the process $\epem\to\ggg$ is fully described
by QED. Measurements of the differential cross-sections from the
four LEP experiments are compared to the QED expectation and limits 
are set on parameters describing physics beyond the Standard 
Model. Three-photon events are used for a direct search for
a photonically decaying resonance produced together with a photon.
}

\section{Introduction}

The process $\epem\to\ggg$, called multi-photon production,
is one of the few processes in high energy $\epem$ scattering
which can be
described by QED only. Since the only free parameter $\alpha(0)$
is precisely measured\footnote{Since the photons are in the final 
state the relevant momentum transfer for the fine-structure constant
is the mass of the photon which is zero.} the Standard Model
expectation is well known. Any deviation would hint at some
new physics. 
In general such effects can be described by cut-off parameters
or in the framework of effective Lagrangian theory. Effects can for 
example be caused by the t-channel exchange of excited electrons 
or the s-channel exchange of gravitons in models with extra dimensions.
Events with three photons in the final state are used to
search directly for a photonically decaying resonance which is produced
together with a photon. 
Results will be presented from the four LEP experiments 
based on the
full LEP2 statistics, including data taken in 2000.\footnote{ALEPH 
results from data taken in 2000 are not yet available.}

\section{Theory}

The Born-level differential cross-section for the process $\eegg$
in the relativistic limit of lowest order QED is given by~\cite{bg}
\be
\xb = \frac{\alpha^2}{s}\;\frac{1+\cos^2{\theta} }{1-\cos^2{\theta} }
    \; ,
\label{eq:born}
\ee
where $s$ denotes the square of the centre-of-mass energy and 
$\theta$ is the scattering angle.
Since the two photons are identical particles, the event angle is 
defined by convention such that $\ct$ is positive.

Possible deviations from the QED cross-section can be parametrised in 
terms of cut-off parameters $\Lambda_\pm$ which correspond to an 
additional exponential term to the Coulomb field~\cite{drell}
as given in Eq. \ref{eq:lambda}.
Alternatively, in terms of effective Lagrangian theory,~\cite{eboli} 
the cross-section depends on the mass scales (e.g. $\Lambda '$) for 
$\rm ee\gamma\gamma$ contact interactions or non-standard 
$\epem\gamma$ couplings. 
The resulting cross-sections are of two general types, either an
angular independent offset to the cross-section or similar to the
form given by Eq. \ref{eq:lambda}.
\be
\xl   =  \xb \pm \; \frac{\alpha^2 s}{2\Lambda_\pm^4}(1+\cos^2{\theta}) 
\label{eq:lambda}
\ee

Recent theories have pointed out that the graviton 
might propagate in a higher-dimensional space where additional 
dimensions are compactified while other Standard Model particles are 
confined to the usual 3+1 space-time dimensions. 
The resulting large number of Kaluza-Klein excitations could be
exchanged in the s-channel of $\eegg$ scattering.
This leads to a differential cross-section~\cite{ad}
 depending on the mass scale $M_{\rm S}$ which should be 
of order of the electroweak scale (${\cal O}(10^{2-3}$GeV))
and a parameter $\lambda$ which is of ${\cal O}(1)$. Ignoring
${\cal O}(M_{\rm S}^{-8})$ terms
$M_{\rm S}^{-4} = \alpha\pi\Lambda^{-4}_\pm$ for $\lambda = \mp 1$.

The existence of an excited electron ${\rm e}^*$ with an
${\rm e}^*{\rm e}\gamma$ coupling would contribute to the photon 
production process via $t$-channel exchange. The resulting 
cross-section depends on the ${\rm e^*}$ mass $M_{\rm e^*}$ and the 
coupling constant $\kappa$ of the $\mathrm{e^* e\gamma}$ vertex 
relative to the $\mathrm{ee\gamma}$ vertex.~\cite{litke}
For large masses $M_{\rm e^*}\approx\sqrt{\kappa}\Lambda_+$.

\section{Radiative corrections}
All cross-sections discussed above are calculated to
${\cal O}(\alpha^2)$.
For higher order QED predictions an exact ${\cal O}(\alpha^3)$ 
Monte Carlo~\cite{mcgen} and a Monte Carlo~\cite{fgam} for 
\mbox{$\epem \to \gamma\gamma\gamma\gamma$} in the relativistic limit 
are available,
but no full ${\cal O}(\alpha^4)$ calculation.
To keep theoretical uncertainties from higher orders below 1\% it is
essential to minimise third order corrections. 
This can be achieved via a proper choice of the scattering angle. 
Whereas in lowest order there are
exactly two photons with the same scattering angle 
\mbox{($|\ct_1| = |\ct_2|$)}
for a measured event the two highest energy photons are in general
not back-to-back ($|\ct_1| \neq |\ct_2|$). This leads to various 
possibilities for the definition of the scattering angle of the event.

The simplest quantity is the average \mbox{$\ct_{\rm av} = (|\ct_1|+|\ct_2|)/2$} which
leads to corrections of up to 30\% at angles of $\ct_{\rm av} \approx 0$.
These large corrections arise since the average does not change if
one photon flips from one hemisphere to the other. This problem can be
avoided using the difference \mbox{$\ct_{\rm dif} = (|\ct_1 - \ct_2|)/2$},
which shows otherwise the same behaviour with corrections up to
10 (15)\% for large values of $\ct_{\rm dif} >$ 0.88 (0.95). 
A physics motivated definition is
the angle in the centre-of-mass system of the two highest energy photons
\mbox{$\cte = |\sin{\frac{\theta_1 - \theta_2}{2}}| \; /
   \; ( {\sin{\frac{\theta_1 + \theta_2}{2}}}) $}.
This definition leads to the smallest corrections of 3-7\% within
the studied angular range of $\cte <0.97$ and is therefore chosen for the 
analyses.

\section{Selection}

The selection of multi-photonic events relies on the photon detection
in the electromagnetic calorimeter ECAL. The ECAL signature however is
the same for $\ggg$ and $\epem(\gamma)$ events. Since the Bhabha
cross-section is huge this background must be suppressed by 4-5 orders
of magnitude. To reject Bhabhas the tracking detectors 
are used to distinguish electrons from photons. This can be difficult 
at small scattering angles where the electrons do not travel the full extent 
of the tracking chamber and two particles can easily be 
reconstructed as one track. Also photon conversions at a small distance 
to the interaction point, i.e. before the first active detector layer
are hard to separate. Since the conversion rate depends on the 
material in the detector the optimal angular range for the selection 
strongly depends on the experiment.
The acceptance ranges given in Tab. \ref{tab:eff} reflect also the 
different angular coverage of the ECAL. 
A very dangerous background is caused by low $s'$ Bhabhas, with
an invariant photon mass just above the threshold of 1 MeV. 
They have the same signature as a photonic event with early conversion
and are badly simulated, since most Bhabha Monte Carlos impose much 
higher cuts on $s'$.

\begin{table}[h]
\caption{Acceptance range and efficiency $\epsilon$ within this 
acceptance range of the four experiments. The efficiency might 
depend slightly on $\sqrt{s}$. 
The assumed systematic error on the efficiency
$\delta\epsilon$ and the radiative corrections $\delta\rho$
are also given. Preliminary L3 results do not include systematic errors.
Other systematic errors are small.
\label{tab:eff}}
\begin{center}
\begin{tabular}{|l|cccc|}
\hline
 & $\ct$ range & $\epsilon$ & $\delta\epsilon$ & $\delta\rho$ \\\hline
ALEPH  & $[ 0, 0.95 ]$ & 83\% & 1.3\% & 1.0\% \\[0.5ex]
 & $[0.035 , 0.731] \cup $ & & & \\[-1ex]
\raisebox{1.2ex}{DELPHI} & $[0.819 , 0.906 ]$ & 
\raisebox{1.2ex}{76\%} & \raisebox{1.2ex}{2.5\%} &
\raisebox{1.2ex}{0.5\%} \\[0.5ex]
L3     & $[ 0, 0.961 ]$ & 64\% & 1.2\%/-- & -- \\[0.5ex]
OPAL   & $[ 0 , 0.90 ]$ & 92\% & 1.0\% & 1.0\% \\
\hline
\end{tabular}
\end{center}
\end{table}

\section{Cross-section results}

\begin{figure}[t]
\epsfxsize=\textwidth
\centerline{\epsfbox{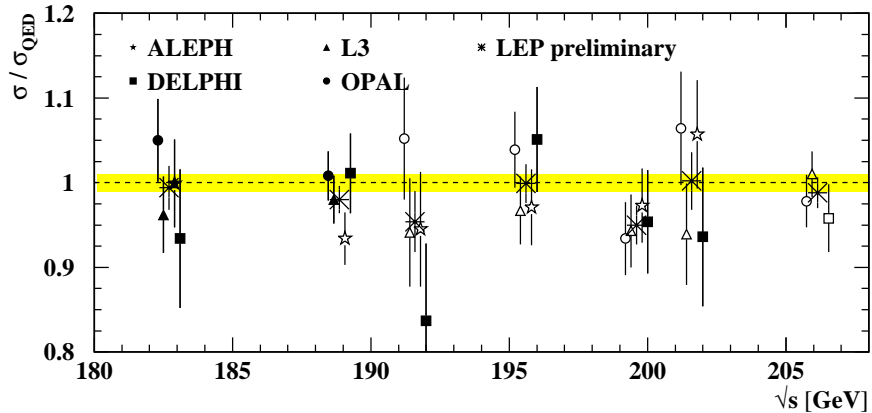}} 
\caption{Measured total cross-section relative to the QED expectation.
The values of single experiments are shown displaced in $\sqrt{s}$
for clarity. Filled symbols represent final results, open symbols 
are preliminary. The error on preliminary L3 cross-sections is 
statistical only. The theoretical error of 1\% is shown as 
shaded band and is not included in the experimental errors. 
  \label{fig:totx}}
\end{figure}

The total and differential cross-sections are measured within the 
angular ranges given in Tab. \ref{tab:eff}. Figure \ref{fig:totx}
shows the total cross-sections normalised to the QED expectation for 
all four LEP experiments \cite{aleph,delphi,l3,opal} 
and their combination. Apart from the common 
theoretical uncertainty the correlated systematic error between 
experiments is negligible. In general there is a very good agreement.
The average over all energies and experiments   
yields $0.980 \pm 0.008 \pm 0.007$ where the first error is
statistical and the second systematic. This is two standard deviations
low, not accounting for the assumed theoretical error
of 1\% which is of the same size as the experimental error.
 
The angular distributions are compared to the differential cross-sections
predicted by various models. No significant deviation from QED was found
and limits given in Tab. \ref{tab:lim} are derived.
For all experiments the limit on $\Lambda_+$ is larger than the
limit on $\Lambda_-$. This effect is not significant 
yet it implies that the observed cross-section in the central region 
of the detectors is smaller than expected.
Small scattering angles are less sensitive to
these limits though their number of events is largest.

\begin{table}[bt]
\begin{center}
\caption{Limits derived from fits to the angular distribution:
the cut-off parameter $\Lambda_\pm$, the mass scale for 
$\rm ee\gamma\gamma$ contact interaction $\Lambda '$, the 
mass of an excited electron $\rm M_{\rm e^\ast}$ and the
mass scale for extra dimensions M$_{\rm S}$ for $\lambda=\pm 1$.
\label{tab:lim}}
\begin{tabular}{|l|cccccc|}
\hline
 \multicolumn{5}{|c}{} & \multicolumn{2}{c|}{M$_{\rm S}$} \\[-1ex]
 [GeV] & $\Lambda_+$ & $\Lambda_-$ & $\Lambda '$ &
    M$_{\rm e^\ast}$ & \small$\lambda$=+1 & \small$\lambda$=--1 \\\hline
 ALEPH  & 319 & 317 & 705 & 337 & 810 & 820 \\
 DELPHI & 354 & 324 & --  & 339 & 832 & 911 \\
 L3     & 385 & 325 & 810 & 325 & 835 & 990 \\
 OPAL   & 344 & 325 & 763 & 354 & 833 & 887 \\
\hline
\end{tabular}
\end{center}
\end{table}

\section{Resonance production}

\begin{figure}[t]
\epsfxsize=10cm 
\centerline{\epsfbox{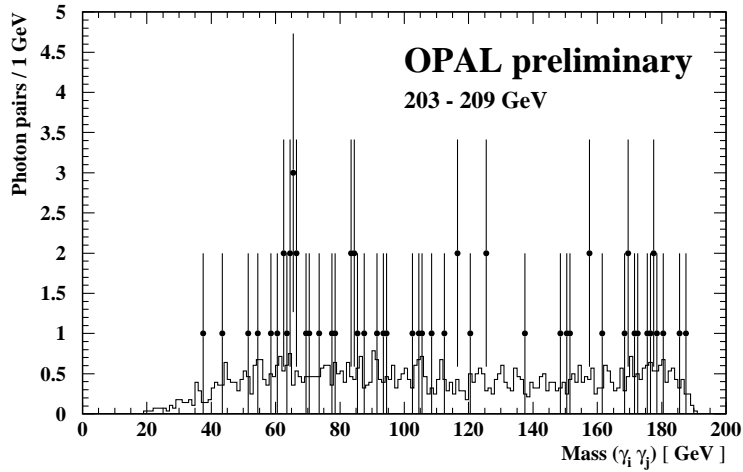} }
\caption{Invariant mass of photon pairs from events with three photons
in the final state. There are three combinations per event. The points
represent the data taken in 2000 by OPAL and the histogram the corresponding
QED expectation. The mass resolution is about 0.5 GeV.
The mass range is limited not only by the centre-of-mass energy but also 
by the imposed cut on the opening angle between photons.
  \label{fig:mass}}
\end{figure}

Three photon final states can originate from a photonically decaying 
resonance $\rm X\to\gamma\gamma$, which is produced together with a 
photon via $\epem\rm\to X \gamma$. Fig. \ref{fig:mass} shows the 
invariant mass of photon pairs in events with exactly three photons.
Within the small statistics the mass distribution is in good agreement
with the expectation from the QED process.

If the Higgs is assumed to be this resonance X
the Standard Model coupling of $\rm H\to\gamma\gamma$
via loops of charged, massive particles is too small to lead to an
observable effect. 
For the Standard Model Higgs the maximum of the branching ratio
for $\rm H\to\gamma\gamma$ is $2.6 \cdot 10^{-3}$ at a Higgs mass of
about 125 GeV and a total Higgs width of $\sim 4$ MeV. 
For larger Higgs masses the branching ratio
decreases due to the increasing $\rm H \to W^+W^-$ contribution.
However, for limits on anomalous couplings in the case of fermiophobic 
Higgs models the
three photon final state gives information which is complementary  
to $\epem\rm\to H Z$ with $\rm H \to\gamma\gamma$ or 
$\epem\rm\to H \gamma$ with $\rm H \to \bar{b}b$.

Anomalous $\rm H\to\gamma\gamma$ couplings can be described 
by~\cite{hagiwara}
\be
{\cal L}_{\rm eff}^{\rm H\gamma\gamma} = 
- \frac{g M_{\rm W}}{\Lambda^2}\frac{\sin^2{\theta_{\rm W}}}{2}
(f_{\rm BB} + f_{\rm WW} - f_{\rm BW} ) {\rm H A_{\mu\nu} A^{\mu\nu}}
\; , \ee
where $\Lambda$ is the energy scale and the three possible
couplings are $f_{\rm BB}$, $f_{\rm WW}$ and $f_{\rm BW}$.
Limits on $\rm \gamma Z$ interaction set strong constraints on 
$f_{\rm BW}$ \cite{bw} which is therefore set to zero. The two other 
parameters are in general assumed to be identical 
$f_{\rm BB} = f_{\rm WW} \equiv F$.

\begin{figure}[t]
\epsfxsize=6cm 
\mbox{\epsfbox{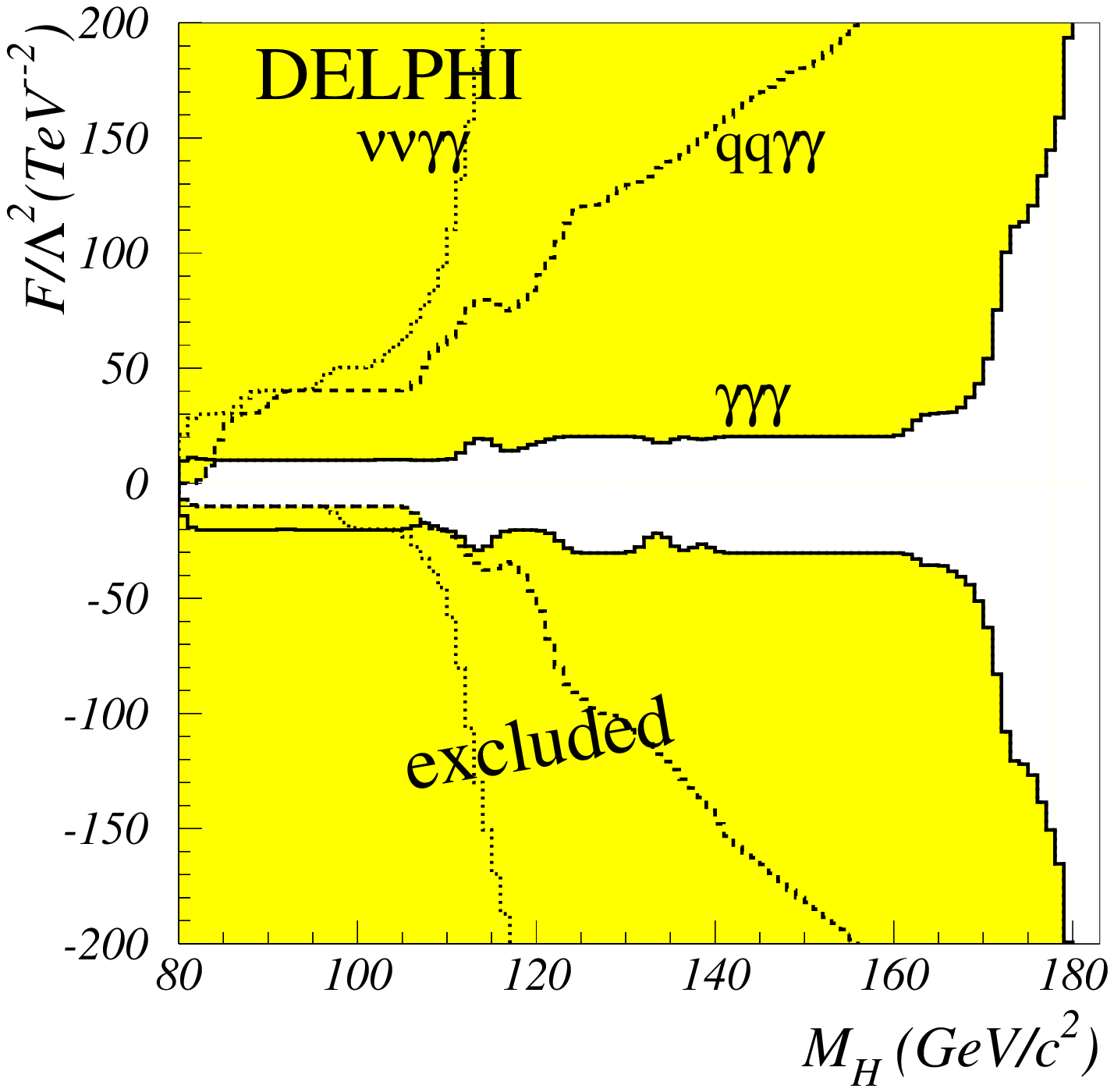}} \hfill
\epsfxsize=6cm 
\mbox{\epsfbox{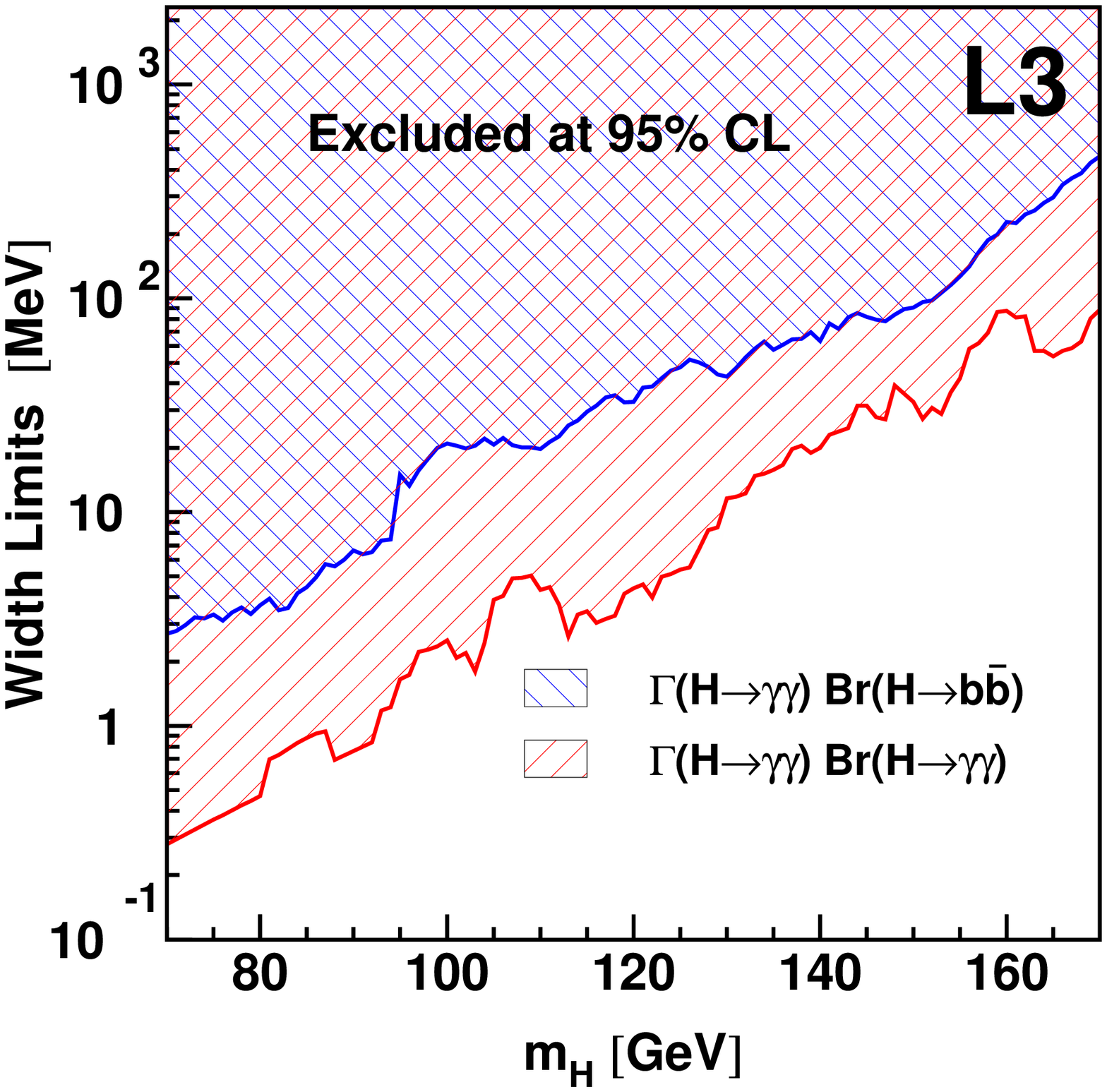} }
\caption{Limits on anomalous $\rm H \to \gamma\gamma$ coupling
depending on the Higgs mass. The DELPHI plot shows limits on 
$F/\Lambda^2$ derived from $\rm H\gamma$ and $\rm H Z$ production.
The L3 plot shows limits on the partial width $\rm H\to\gamma\gamma$
from $\epem\rm\to H\gamma$ with 
$\rm H\to b\bar{b} \mbox{ and } \gamma\gamma$. 
Data taken until 1999 were used.
  \label{fig:higgs}}
\end{figure}

In Fig. \ref{fig:higgs} the limit \cite{delhiggs} on $F/\Lambda^2$ 
is shown from $\rm H\to\gamma\gamma$ decay
for the two processes $\epem \rm\to H \gamma$ and $\epem\rm \to H Z$, 
with $\rm Z \to q\bar{q} \mbox{ and } \nu\bar{\nu}$. Although
$\rm HZ$ production has the higher sensitivity for $F/\Lambda^2$ at low
Higgs masses, $\rm H\gamma$ production provides strong limits up to
$M_{\rm H}\sim$ 170 GeV. 

The partial width of $\rm H\to\gamma\gamma$ 
is studied with the process $\epem\rm\to H \gamma$. 
Fig.~\ref{fig:higgs} shows limits \cite{l3higgs} 
from three photon final states
($\rm H\to \gamma\gamma$) which are stronger than those obtained
from $\rm b\bar{b}\gamma$ events ($\rm H\to b\bar{b}$). 

\section{Conclusion}
The process $\eeggg$ provides high statistics data for the test of the 
Standard Model. Combining all four LEP experiments  a
precision of 1\% for the total cross-section is reached. 
Since this process is dominated by QED a precise prediction is in 
principle possible. However, since calculations are available only up 
to next-to-leading order a theoretical error of about 1\% has to be 
taken into account while searching for deviations from the 
Standard Model. The observed total cross-section is two standard deviations
below the expectation not accounting for this theoretical error.
Three photon final states are interesting for the search for photonically 
decaying resonances, which are produced along with a photon.
The Standard Model prediction for the $\rm H\to\gamma\gamma$ branching ratio
is too small to lead to an observable cross-section at LEP
hence limits on anomalous couplings are placed.

\end{document}